\documentclass[aps,prl,reprint,twocolumn,amsmath,amssymb,floatfix]{revtex4-1}

\usepackage[unicode=true,pdfusetitle,bookmarks=false,breaklinks=true,pdfborder={0 0 0},backref=false,colorlinks=true]{hyperref}
\hypersetup{linkcolor=blue,urlcolor=blue,citecolor=blue}

\usepackage{graphicx}

\newcommand{\kb}{{k_B}}

\newcommand{\Tcquasi}{T_{\rm qc}}

\begin{document}

\title{Two-dimensional spin-imbalanced Fermi gases at non-zero temperature:\\
Phase separation of a non-condensate}

\author{Chien-Te Wu}
\author{Brandon M. Anderson}
\author{Rufus Boyack}
\author{K. Levin}
\affiliation{James Franck Institute, University of Chicago, Chicago, Illinois 60637, USA}

\begin{abstract}
We study a trapped two-dimensional spin-imbalanced Fermi gas over a range of temperatures.
In the moderate temperature regime, associated with current experiments, 
we find reasonable semi-quantitative agreement with the measured
density profiles 
as functions of varying spin imbalance and interaction strength.
Our calculations show that, 
in contrast to the three-dimensional case, the phase separation which
appears as a spin balanced core, 
can be associated with non-condensed fermion pairs.
We present predictions at lower temperatures where a quasi-condensate will 
first appear, based on the  
pair momentum distribution and following the protocols
of Jochim and collaborators.
While these profiles also indicate phase separation, they exhibit
distinctive
features which may aid in identifying the condensation regime. 
\end{abstract}

\maketitle

Ultracold Fermi gases are a valuable resource
for learning about highly correlated superfluids.
Their utility comes from their tunability~\cite{KetterleVarenna}
which allows the dimensionality, band structure, interaction
strength, and spin imbalance to be freely varied.
With these various parameters one can, in principle, simulate
a number of important condensed matter systems ranging
from preformed pair and related
effects in the high $T_c$ cuprates~\cite{ourreview,Strinatispectral,Loktev}
to intrinsic topological superfluids~\cite{Zhaireview,dasSarma1,Iskin} 
and other exotic pairing states.

In this paper we focus on recent experiments~\cite{Thomas2D,Bakr} of two-dimensional (2D) spin-imbalanced Fermi gases. 
These address the interesting conflict between the tendency towards enhanced pairing 
(which is associated with two dimensionality~\cite{Randeria2d}),
and spin imbalance which acts to greatly undermine pairing.
These imbalance effects  are believed~\cite{Wilczek} to have related effects in studies of color
superconductivity and quark-gluon plasmas. 
In condensed matter systems, lower dimensional imbalanced superfluids are thought to be ideal for
observing more exotic phases, such as the elusive LOFF state~\cite{LOFF_Review}, or algebraic order~\cite{Kosterlitz,Berezinskii}.

The approach we use has been rather successful in addressing
2D low temperature quasi-condensation~\cite{our2D}
in balanced gases. In this paper we present
predictions for future very low temperature experiments on spin-imbalanced gases.
Importantly, our calculations, which find no true long range
order, are consistent with the
Mermin-Wagner theorem~\cite{MW}. Following the
experiments of Jochim and collaborators~\cite{Jochim1,Jochim2}, we
show how quasi-condensation is reflected
in the pair-momentum distribution which has a strong
peak at low momentum. This peak, which we study throughout the
crossover from BCS to BEC,  disappears somewhat abruptly
at a fixed temperature, $T_{\rm qc}$, which denotes quasi-condensation.
We find  $T_{\rm qc}$ varies only weakly with the polarization.

A central finding in this paper is that
2D spin-imbalanced systems in a trap exhibit a new form
of phase separation involving non-condensed pairs 
appearing primarily in the center region of the trap.
This is to be contrasted with 3D gases~\cite{PLKLH06,Shin}
where the phase separation is associated with a true condensate.
In the 2D polarized case, because almost all the pairs reside in
the central portion of the trap, 
this leads to a nearly balanced core, as observed in
recent experiments~\cite{Thomas2D,Bakr}.
As one goes to larger radii, there are one (at
low $T$) or two (at moderate $T$) additional shells.
The outermost shell is to be associated with a Fermi gas of
majority atoms. An intermediate shell (if it exists) is partially
polarized and consists
of broken pairs with majority and some minority atoms. 
Our calculations indicate a necessary but not sufficient condition
for quasi-condensation is that a partially polarized 
intermediate shell will not be present.
Instead, there is an abrupt transition from a balanced core to a normal
Fermi gas.


There is a sizeable literature on the 
mutual effects of imbalance and two dimensionality in Fermi superfluids.
Experiments have focused on a polaronic
interpretation of radio frequency~\cite{Thomaspolaron}
and thermodynamic data. Theory has focused either on the very low temperature
region (both the ground state~\cite{Heathground,Conduitground} and Kosterlitz
Thouless regimes~\cite{KliminBKT,KliminBKT2}), 
on possible LOFF phases ~\cite{Sheehy}
and on the polaronic limit 
\cite{Parishpolaron,Parishpolaron2}
where the spin imbalance is extreme. 
Here we consider the entire range of temperatures 
and polarizations, where there are few theoretical studies~\cite{MonteCarlo,Caldas}
using a theory~\cite{our2D} which is consistent with earlier ground state work~\cite{Heathground}
and with recent experimental studies~\cite{Jochim1,Jochim2}  of 2D balanced gases.
We also present a rather successful comparison
with experiments~\cite{Thomas2D} performed at moderate temperatures
for the in-situ density profiles across the range of BCS to BEC.

\textit{Theoretical formalism.---}
The present theory is based on the BCS-Leggett ground state~\cite{Leggett}
generalized to include polarization effects~\cite{Hethermo,ChienRapid,Chien06,ChenHeChien}
and to higher temperatures. Without showing the details, which have
appeared in the recent literature~\cite{our2D}, we present
two coupled equations that define a self-consistent fluctuation theory:
\begin{align}
\label{eq:1}
\sum_{\mathbf{k}}\left[\frac{1-f\left(E_{\mathbf{k}\uparrow}\right)-f\left(E_{\mathbf{
k}\downarrow}\right)}{2E_{\mathbf{k}}}-\frac{1}{2\epsilon_{\mathbf{k}}+\epsilon_{B}}
\right]&=a_{0}\mu_{{\rm pair}},\\ 
\label{eq:2}
\sum_{\mathbf{q}}b\left(\frac{\mathbf{q}^{2}}{2M_B}-\mu_{{\rm pair}}\right) &=a_{0}\Delta^{2}.
\end{align}
%
Here, the two-band Bogoliubov quasiparticle dispersion $E_{\bf k\sigma} = \sigma \, h + \sqrt{\xi_{\bf k}^2 + \Delta^2}$ ($\uparrow$, $\downarrow$ correspond to $\sigma=+1,-1$ respectively) is constructed from the bare fermions with excitation spectrum $\xi_{\bf k} = \epsilon_{\bf k} - \mu$ and pairing gap $\Delta$. The fermions of mass $m$ and momentum ${\bf k} = (k_x,k_y)$ have a single particle excitation spectrum $\epsilon_{\bf k} = \mathbf{k}^2/2m$, a fermionic chemical potential $\mu$. An effective Zeeman field $h>0$ shifts the energy of (majority) spin-up relative to the (minority) spin-down Fermi surfaces. We have also introduced the usual Bose and Fermi distribution functions $b(x)$ and $f(x)$, and included the two-particle binding energy $\epsilon_B$~\cite{Loktev} to regularize Eq.~(\ref{eq:1}). Throughout this paper we set $\hbar=1$.

\begin{figure}[ht] 
\includegraphics[width=3.2in,clip]{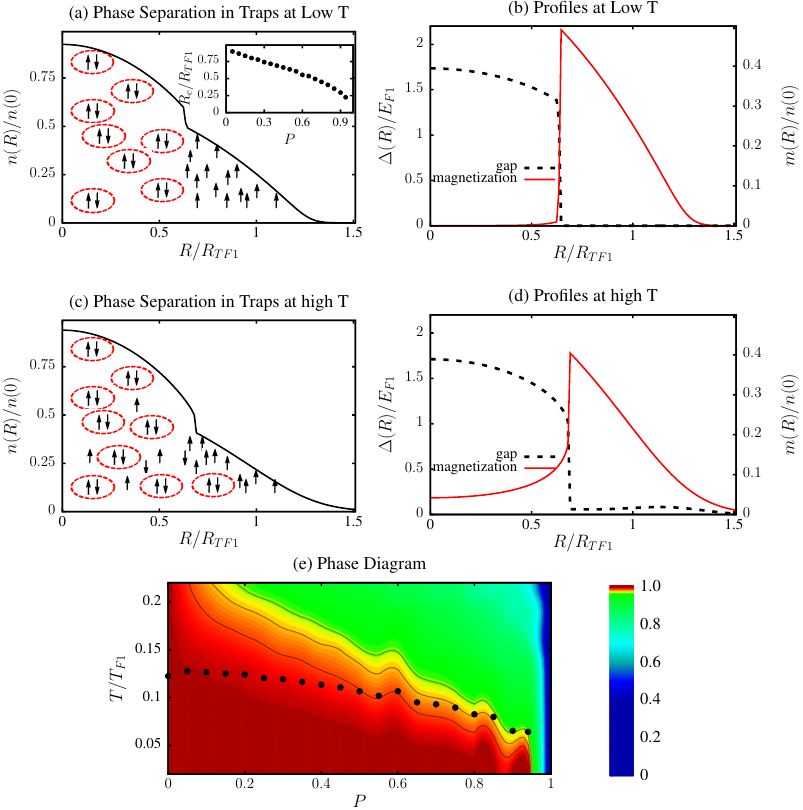}
\caption{
This figure contrasts the nature of phase separation in a harmonic trap
at low temperatures ((a) and (b)) where quasi-condensation occurs
($T/T_{F1}=0.06$)
and moderate temperatures ((c) and (d))
($T/T_{F1}=0.22$) more appropriate to experiments~\cite{Thomas2D,Bakr};
the binding energy is fixed at $E_{F1}/\epsilon_B=0.75$.
Black lines represent the local density $n(R)$ in (a) and (c), while
in (b) and (d) the pairing gap $\Delta(R)$ is black (dashed) and the magnetization $m(R)$ is red.
Panel (a) shows the ``two shell" structure:
the core region, next to a fully polarized region,
is occupied only by pairs.
The radius at which the gap turns off abruptly at low $T$ is
indicated as an inset in (a).
The density profile in panel (c) exhibits a ``three shell''
structure: the almost balanced core region is followed
by a transition region that is partially polarized and
the edge is fully polarized.
Finally panel (e) presents
a phase diagram where the color contours indicate the central
balance ratio, $\tilde{p}(0) = n_\downarrow(0)/n_\uparrow(0)$, 
of minority to majority atoms at the trap center.
The three contours mark values of $99\%$, $98\%$, and $97\%$ for this ratio.
The black dots mark the onset of quasi-condensation, as defined in Eq.~(\ref{eq:4}).
}
\label{fig:1}
\end{figure}

The key physics in our system is captured by Eqs.~(\ref{eq:1})-(\ref{eq:2}),
which reflect the natural equilibrium between 
fermionic quasiparticles and non-condensed pairs or bosons. 
Specifically, Eq.~(\ref{eq:2}) introduces non-condensed bosonic degrees
of freedom which have momentum ${\bf q}$, mass $M_B$, and chemical potential $\mu_{\rm pair}$.
($M_B$ and the constant $a_0$ are calculated from an expansion of
a $t$ matrix describing paired bosons. See the Supplemental Material in Ref.~\cite{our2D}
for a precise definition.)

These fluctuations are not present in the strict mean-field theory of BCS;
if one sets the pair chemical potential $\mu_{\rm{pair}}$ to
zero, then Eq.~(\ref{eq:1}) reduces to the usual mean-field
equation for a polarized gas, specifying the gap parameter $\Delta$.
Including these fluctuations then allows one to solve for the two unknowns:
$\Delta$ and $\mu_{\rm pair}$. 
The fermions are associated with the energy
$\Delta$ needed to break apart pairs,
and the non-condensed bosonic pairs have a
self-consistently determined
chemical potential $\mu_{\rm pair}$, which depends on the pairing
gap $\Delta$. Here the number density of pairs (bosonic number)
is given by $n_B = a_0 \Delta^2$. 
The more non-condensed bosons which are present, the
larger the pairing gap.
That these bosonic degrees of freedom cannot condense in
2D allows us to incorporate the constraints of the 
Mermin-Wagner theorem.
We can think of these as the introduction of fluctuation effects.

\begin{figure}
\includegraphics[width=3.3in,clip]{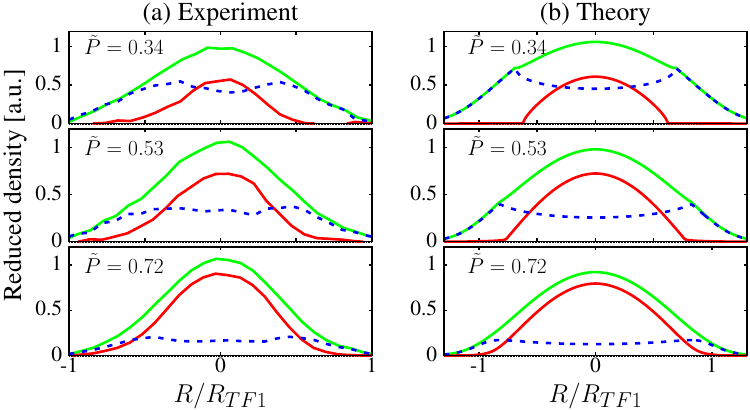}
\caption{
Comparison of integrated column density profiles of (a) experiment~\cite{Thomas2D} and (b) theory
for a trapped system with $E_{F1}/\epsilon_B=0.75$ and $T/T_{F1} = 0.22$.
The green, red, and blue curves are the reduced densities (see~\cite{Thomas2D} for definition)
of the majority, minority, and magnetization (difference), respectively.
The legend indicates the total balance ratio $\tilde{P}=N_{\downarrow}/N_{\uparrow}$.
A transition to a nearly balanced core is seen in both theory and experiment.}
\label{fig:2}\end{figure}

\begin{figure}
\includegraphics[width=3.3in,clip]
{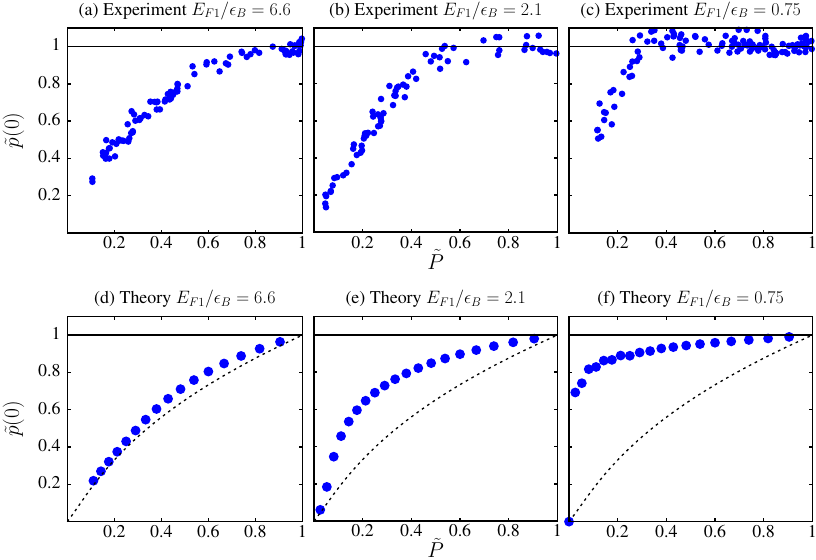}
\caption{Comparison of theoretical and experimentally measured~\cite{Thomas2D} values of the central balance ratio $\tilde{p}(0)=n_\downarrow(0)/n_\uparrow(0)$, 
at $T/T_{F1}=0.22$.
The dashed curves give the ideal Fermi gas limit ($\epsilon_B=0$);
the solid black curves are guides to the eye.}
\label{fig:3}\end{figure}

\begin{figure*}
\includegraphics[width=5.5in,clip]{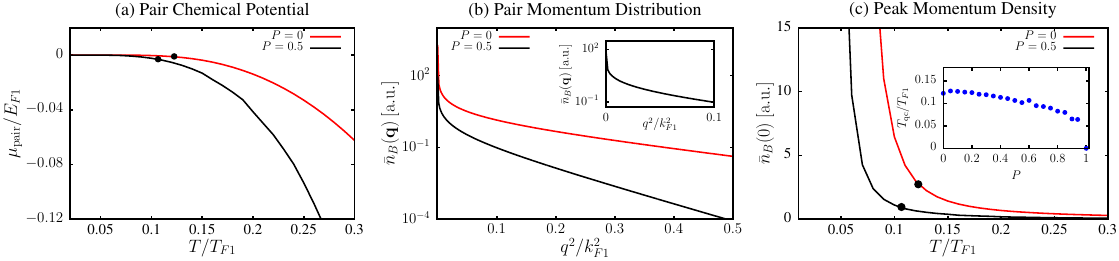}
\caption{
Characteristics of quasi-condensation at a binding energy $E_{F1}/\epsilon_B=0.75$.
Pair chemical potential (a) for a polarized (black, $P=0.5$) and unpolarized (red) Fermi gas.
The small and non-zero size of $\mu_{\rm pair}$ reflects an exponential suppression at low temperatures~\cite{our2D}.
(b) This leads to a low-momentum peak in the pair momentum distribution $n_B({\bf q}=0)$ at low temperatures.
(c) The dependence of the $n_B(0)$ peak on temperature allows the extraction (dots) of $\Tcquasi$ in Eq.~(\ref{eq:4}).}
\label{fig:4}\end{figure*}


In experiments,
the effective Zeeman field
$h$ and total chemical potential $\mu$
derive from a magnetization $m = n_\uparrow-n_\downarrow$
and number density $n=n_\uparrow + n_\downarrow$. Thus we set the
fermionic chemical potentials using
the number equation 
\begin{equation}
n_\sigma=\frac{1}{2}\sum_{\mathbf{k}}\left[
\left(1-\frac{\xi_{\mathbf{k}}}{E_{\mathbf{k}}}\right)f\left(-E_{\mathbf{k}\bar\sigma}\right)+
\left(1+\frac{\xi_{\mathbf{k}}}{E_{\mathbf{k}}}\right) f\left(E_{\mathbf{k}\sigma}\right)
\right]
\end{equation}
for the number density of species $\sigma=-\bar{\sigma}$.

To account for the trapping potential, we apply the local density approximation (LDA) to a system 
with total atom number $N_\uparrow$ $(N_\downarrow)$ of majority (minority) carrier.
Here we replace $\mu\to\mu(R)\equiv \mu_0 -\frac{1}{2}m\omega^2 R^2$,
and $\Delta\to\Delta(R)$, where $\mu_0$, $\omega$, 
and $R$ represent the central fermionic chemical potential,
the trap frequency, and position respectively. 
Derived quantities such as the magnetization $m(R)$, number density $n(R)$, and pair mass $M_B(R)$ gain local dependence.
However, the effective Zeeman field is homogeneous throughout the trap.
Where relevant, we express energy and local position in units of the majority spin Fermi energy, 
$E_{F1}=\omega\sqrt{2N_\uparrow}$, and Thomas-Fermi radius $R_{TF1} = \sqrt{2E_{F1} / m\omega^2}$ respectively;
we take $\omega/E_{F1}=1/40$ comparable to Ref.~\cite{Thomas2D}.

In what follows, it will be convenient to define a local polarization $p(R)=m(R)/n(R)$, a total polarization $P=\left(N_\uparrow-N_\downarrow\right)/\left(N_\uparrow+N_\downarrow\right)$. To connect with recent experiments~\cite{Thomas2D}, we will also define a ``balance ratio'' $\tilde{p}(R)=n_\downarrow(R)/n_\uparrow(R)$, and similarly for a total balance ratio $\tilde{P}=N_\downarrow/N_\uparrow.$

\textit{Numerical results.---}
Figure~\ref{fig:1} serves to clarify the concept of phase separation
in a trapped 2D gas for both low ($T/T_{F1} = 0.06$)
and moderate temperatures ($T/T_{F1}=0.22$) regimes.
The former are applicable to the quasi-condensation regime discussed below,
while the latter are closer to the regimes
studied experimentally~\cite{Thomas2D,Bakr}.
We consider an intermediate binding energy $E_{F1}/\epsilon_B=0.75$.
The density profiles as a function of position
in these two temperature regimes are plotted in
panels (a) and (c), along with a cartoon illustration of the
nature of the gas, as the radius changes. 
Panels (b) and (d) provide useful information on the
gap profiles (black dashed) and magnetization profiles (red).
The radius at which the gap turns off abruptly at low $T$ is
indicated as an inset in (a).

At the lower temperatures there is
an abrupt boundary separating a fully paired state in the core (indicated
by the paired spins in the cartoon) and a non interacting fully polarized
gas of majority spins (also represented in a cartoon fashion).
We refer to these profiles as containing only ``two shells": composite bosons
at the core and majority fermions surrounding it.  Importantly, one sees that the
magnetization in Fig.~\ref{fig:1}(b) and the gap both change nearly discontinuously.

Although the number density profile in Fig.~\ref{fig:1}(c) 
behaves similarly to its low-$T$ counterpart,
one sees here (using information about the calculated gap, local polarization, and magnetization), that
there are now ``three shells'' in the structure, as shown
in the cartoon. The core region contains mostly 
non-condensed pairs with very little magnetization. As one goes away 
from the trap center, the local magnetization initally increases resulting in a central shell. 
Finally, toward the edge of the trap where the magnetization drops, 
the gas is non-interacting ($\Delta(R) = 0$) and fully polarized.

These calculations suggest that the magnetization versus position $R$ serves
as a kind of thermometry.
In particular, that we are able to
associate the lower-$T$ behavior with the existence of a 
(quasi-)condensate, can be inferred from the
phase diagram plotted in Fig.~\ref{fig:1}(e).
Here the vertical and horizontal axes
represent temperature $T$, and total polarization $P$, respectively.
As indicated in the legend, the colors more precisely
correspond to the ratio of
minority to majority spins at the trap center.
The three contours mark $99\%$, $98\%$, and $97\%$ for
the ratio.
The black dots on the phase diagram show where we find pair
(quasi-)condensation, as will be discussed below.
It should be clear that this quasi-condensate essentially always
appears in conjunction with our ``two shell" profiles.


\textit{Comparison with Experiment.---}
The general picture described above has  
consequences that can be directly compared to recent experiments. 
In Fig.~\ref{fig:2}, as in experiment~\cite{Thomas2D},
we plot ``column density'' profiles 
for majority and minority components in the trap along with the
difference profile (local magnetization), for three different values of
the balance ratio $\tilde{P}$.
(The total polarization increases as one goes upward on the three panels).
The counterpart experimental data is plotted
on the left along with theory curves on the right.
In the calculations, we consider
fixed moderate temperature $T/T_{F1}=0.22$.

This observation of phase separation of a non-condensate underlines some of the
same points as in 
the 2D density profiles
shown in Fig.~\ref{fig:1}. Here, however, one sees how this physics is reflected
in actual experiments.
Indeed, there is a particularly interesting indicator of this form of
phase separation.
The ratios of minority to majority 2D densities at the trap center
have been measured by Thomas and collaborators~\cite{Thomas2D}.
These experiments investigate the variation as one
crosses from more BCS to more BEC like behavior.
They observe (see Fig.~\ref{fig:3}, top panels)
the somewhat striking result that, away from the
BCS regime, there is a rather abrupt transition from
a balanced core to an unpaired phase at a critical polarization
(which is presumably temperature dependent).
The constancy of the data points
indicates the very strong tendency to maintain
maximal pairing until it is no longer possible. The abrupt
drop occurs presumably
because one has crossed the so-called $T^*(P)$ line.
This temperature $T^*$ marks the end-point of a pairing gap,
often called the pseudogap.

In Fig.~\ref{fig:3} we present a comparison between theoretical and experimental results,
plotting the central balance ratio $\tilde{p}(0)$ as a function of the total balance ratio $\tilde{P}$.
In our theoretical analysis we fix the temperature
for all panels at $T/T_{F1}=0.22$. 
In the stronger pairing cases (with
$E_{F1}/{\epsilon_B}=0.75$
and $E_{F1}/{\epsilon_B}=2.1$, 
as shown in the two panels to the right)
the persistence of a balanced core for a range of
total polarizations is observed. The theory curves are not quite as flat as in
experiment, which suggests that theory temperatures are slightly high in comparison;
in both cases the downward departure is reasonably sudden.

These curves reflect simple changes in the trap profile; 
as $\tilde{P}$ increases, the boundary between 
balanced and imbalanced regions moves toward the trap center,
(shown in the inset to Fig.~\ref{fig:1}(a))
while not affecting
the magnetization at the precise ``center''. 
For smaller $\epsilon_B$ at finite $T$, the balanced core
is narrower and the magnetization at the center also increases  
more rapidly as compared to larger $\epsilon_B$. 
At sufficiently high $P$, for this temperature regime the
system is driven normal and the profiles are those of an
ideal Fermi gas.

\textit{Quasi-condensation at very low temperatures.---}
We turn now to the lowest temperature regime away from
$T\equiv 0$, where true long-range order is not possible,
at least for a homogeneous system in the thermodynamic limit. 
The evidence from experiments~\cite{Jochim1,Jochim2}
on 2D spin-balanced Fermi gases suggests that
the bosonic degrees of freedom (accessed by rapid magnetic
field sweeps) exhibit strong $\left| {\bf q} \right| \rightarrow 0$ peaks 
in their momentum distribution, represented by a trap-average, denoted $\bar{n}_B({\bf q})$,
of $n_B({\bf q})=b\left({{\bf q}^{2}}/{2M_B}-\mu_{{\rm pair}}\right)$ appearing in Eq.~(\ref{eq:2}). 
What is most significant~\cite{our2D} is that
these peaks disappear rather abruptly at a particular temperature,
$T_{\rm qc}$, which one associates with 
the onset of quasi-condensation. 

Following
the same analysis for a spin-imbalanced Fermi gas,
in Fig.~\ref{fig:4}(a) we plot
the pair chemical potential $\mu_{\rm pair}$ at the trap center for
an unpolarized (in red) and a polarized gas (in black, $P=0.5$). We find $\mu_{\rm pair}$ 
serves to determine the size of the peak structure in
$\bar{n}_B({\bf q})$ as can be seen from Eq.~(\ref{eq:2}). In both the balanced and
imbalanced cases, this pair chemical potential is
found to nearly vanish at low temperatures, 
signifying a bosonic momentum distribution that is 
sharply peaked but never acquires a macroscopic condensate. 
Moreover, it is seen that the effects of spin imbalance are relatively minor,
resulting in only a small quantitative shift in $\mu_{\rm pair}$ compared to the balanced case.

Figure~\ref{fig:4}(b) presents the counterpart plots of $\bar{n}_B({\bf q})$ versus ${\bf q}$ where
the two curves correspond to $P=0$ in red and $P=0.5$ in black.
The latter is enlarged in the inset, where the peak
in the momentum distribution at ${\bf q}=0$ is evident.
The temperature dependence of this peak is reflected in Fig.~\ref{fig:4}(c).
The solid dots indicate the temperature, $T_{\rm qc}$,  at which the pair
chemical potential begins to deviate from effectively zero.
Taking the deviation point as in Ref. \onlinecite{our2D} (which roughly
corresponds to about a $1\%$ shift from the background) yields
\begin{equation}
\kb \Tcquasi \approx \frac{\pi}{2.3} \frac{\hbar^2 n_B(T=\Tcquasi)}{M_B(T=\Tcquasi)},
\label{eq:4}
\end{equation}
where we use the Bose number density $n_B(T=\Tcquasi)$ and the pair mass $M_B(T=\Tcquasi)$ at the trap center.
The inset of Fig.~\ref{fig:4}(c) presents a plot of this
quasi-condensation temperature as a function of total polarization. 
The effects of polarization on this
temperature are relatively weak, presumably because of the phase
separated and fully balanced spin core. 

These same results are summarized by the black dots in
Fig.~\ref{fig:1}(e) which presents a generalized
phase diagram indicating the $P$, $T$ parameters at which there
is phase separation, as represented by the imbalance at the trap center.
We note that the characteristic inner-core radius, which is plotted as an inset in Fig.~\ref{fig:1}(a), 
shows that for moderate polarizations the range in radii over which one has pairing
(and therefore quasi-condensation) is restricted.
This makes it difficult to perform the analysis that addresses power laws vs exponential
fitting functions in the Fourier transform of $\bar{n}_B({\bf q})$, 
which was identified~\cite{Jochim2} with $g_1(r)$.
For the unpolarized case, such an analysis~\cite{our2D} further substantiated
the identification of $ T_{\rm qc}$ with the expression in Eq. (\ref{eq:4}).

\textit{Conclusions.---}
A goal of this paper has been to emphasize the distinction between
the paired (normal state) and the lower temperature quasi-condensed phase of a
2D polarized gas. 
We show that both are associated with a balanced or nearly balanced core,
but the nature of the related phase separation is somewhat distinctive,
leading to more abrupt boundaries
when quasi-condensation is present.
Proving the existence or non-existence of true phase coherence
would lead to a significant advance in the understanding
of the physics of 2D Fermi gases.
As in previous work~\cite{Jochim1,Jochim2,our2D} true superfluidity
in 2D has not been established here or in experiments. 
This will require future experimental probes related to coherence features,
including interference measurements,
or detection of the presence of collective modes in Bragg scattering.

\textit{Note Added.---} Near completion of the manuscript we learned
of a recent preprint~\cite{Bakr} which addressed experimentally
many of the same findings as contained in our manuscript.

\textit{Acknowledgments.---} This work was supported by NSF-DMR-MRSEC 1420709.
We thank J. Thomas for stimulating conversations and
for sharing his data. We are grateful to W. Bakr and D. Mitra
for useful discussions and information about their experiment.

\bibliography{Review}

\end{document}